\documentclass[prl,twocolumn,showpacs,preprintnumbers,amsmath,amssymb,superscriptaddress]{revtex4}

\usepackage{graphicx}
\usepackage{dcolumn}
\usepackage{bm}

\begin{document}


\title{Devil's crevasse and macroscopic entanglement in two-component Bose-Einstein condensates}
\author{Tim Byrnes}
\affiliation{National Institute of Informatics, 2-1-2
Hitotsubashi, Chiyoda-ku, Tokyo 101-8430, Japan}

\date{\today}

\begin{abstract}
Spin coherent states are the matter equivalent of optical coherent states, where a large
number of two component particles form a macroscopic state displaying quantum coherence.  Here we give a detailed study of entanglement 
generated between two spin-$1/2$ BECs due to an $ S^z_1 S^z_2 $ interaction. The states that are generated show a remarkably rich structure showing fractal characteristics.  In the limit
of large particle number $N $, the entanglement shows a strong dependence upon whether the entangling
gate times are a rational or irrational multiple of $ \pi/4 $.  We discuss
the robustness of various states under decoherence and show that despite the large number of
particles in a typical BEC, entanglement on a macroscopic scale should be observable as long as the gate times are less than 
$ \hbar/J\sqrt{N} $, where $ J $ is the effective BEC-BEC coupling energy. Such states are anticipated to be useful for various quantum information applications such as quantum teleportation and quantum algorithms. 
\end{abstract}

\pacs{03.67.Lx,67.85.Hj,03.75.Gg}
\maketitle

Spin coherent states have been realized in Bose-Einstein condensates (BECs) with the ability of coherent control \cite{bohi09} and spin squeezing \cite{riedel10,gross12}.  One of the main applications of such states is thought to be quantum metrology, where spin squeezed states are created to reduce 
quantum noise in a particular spin direction, to go beyond the standard quantum limit  \cite{sorensen00}.  Spin squeezing
is itself a kind of entanglement \cite{toth09,gross12}, but is a type that is localized in one spatial location, since it manifests itself as
a correlation between bosons within the same BEC. Other types of entanglement of a non-local variety are also possible, as demonstrated 
in the experiment of Ref. \cite{lettner11}.  Here, a single atom was entangled with magnon states within a BEC via optical fiber at 
two distant locations.  Using similar techniques, methods were proposed for generating remote entanglement between two BECs, such 
that entanglement between two BECs would appear to be within experimental possibility 
\cite{byrnes12,treutlein06b}.  Such entangling gates, typically of the form $ S^z_1 S^z_2 $, form a fundamental interaction that is necessary for quantum information processing, where in analogy to standard qubits, one and two qubit gates
form a universal set \cite{byrnes12}. The $ S^z_1 S^z_2 $ interaction forms the analogue of the CNOT gate between two BECs, a fundamental
gate for many quantum circuits. Tasks such as Deutsch's algorithm \cite{byrnes12b}, Grover's algorithm \cite{byrnes12}, and quantum teleportation using BECs \cite{pyrkov13} also rely on the creation of entanglement using the $ S^z_1 S^z_2 $ interaction, and thus it is of both fundamental and practical interest to understand at a deeper level the types of states that are produced by this interaction.   

In this paper we show that such entangling gates have a surprisingly rich structure in terms of the entanglement generated. In addition to potential quantum information processing applications, it is interesting from a physical perspective as it is one of the rare models
in quantum physics where a fractal structure is produced, which is remarkable considering the relative simplicity of the model.  We also calculate the effects of decoherence on the entangled state and show that a particular set of states that should be robust despite the macroscopic nature of the BECs.  The types of states formed are reminiscent optical coherent states with a self-Kerr interaction, where Schrodinger cat state are formed at particular evolution times \cite{milburn86,stobinska08}.

\paragraph{Entangling BECs}

We define the spin coherent state, suitable for BECs, to be
\begin{align}
\label{singlequbitstate}
|\alpha, \beta \rangle \rangle \equiv \frac{1}{\sqrt{N!}} \left( \alpha a^\dagger + \beta b^\dagger \right)^N |0 \rangle ,
\end{align}
where $ a^\dagger, b^\dagger $ are bosonic creation operators for two orthogonal quantum states, $ \alpha, \beta $ are 
arbitrary complex coefficients such that $ | \alpha |^2 + | \beta |^2 = 1 $, and we consider the total number of atoms
to be fixed to $ N $. Such spins may be manipulated on the Bloch sphere in precisely the same way as standard qubits using Schwinger boson operators $ S^x = a^\dagger b + b^\dagger a, S^y = -i a^\dagger b + i b^\dagger a, S^z = a^\dagger a - b^\dagger b $ \cite{byrnes12}.  This analogy to standard qubits suggests that the two-component BEC
may be thought as a ``BEC qubit'', where the spin coherent state encodes one qubit worth of information. 
Let us now consider two such spin coherent states, and consider an entangling operation $ J S^z_1 S^z_2 $ between the two BEC qubits (Figure 1a). 
Starting from two BEC qubits in maximal $ x $-eigenstates we obtain
\begin{align}
& e^{-i S^z_1 S^z_2 \tau} | \frac{1}{\sqrt{2}}, \frac{1}{\sqrt{2}} \rangle \rangle_1 
| \frac{1}{\sqrt{2}}, \frac{1}{\sqrt{2}} \rangle \rangle_2 \nonumber \\
& = \frac{1}{\sqrt{2^N}} \sum_k  \sqrt{N \choose k} | \frac{e^{i(N-2k)\tau}}{\sqrt{2}} , \frac{e^{-i(N-2k)\tau}}{\sqrt{2}} \rangle \rangle_1 | k \rangle_2 ,
\label{entangledstate}
\end{align}
where $ |k \rangle_i = \frac{(a^\dagger_i)^k (b^\dagger_i)^{N-k}}{\sqrt{k! (N-k)!}} $ are eigenstates of $ S^z $ (Fock states), and time is measured in units of $ \hbar/J $. The subscripts $ i = 1,2 $ label the two BECs.  As a first glimpse into the unusual behavior of the entangled state (\ref{entangledstate}), let us plot the von Neumann entropy
$ E_S = - \mbox{Tr} ( \rho_1 \log_2 \rho_1 ) $ where $ \rho_1 $ is the density matrix of (\ref{entangledstate}) with the partial trace over BEC qubit 2 taken. Figure 1a shows the entanglement generated between the two BEC qubits, comparing to the standard qubit case ($ N= 1$) and for BEC qubits with $ N = 100$. For $ N = 1 $ we see the expected behavior, where a maximally entangled state is reached at $ \tau = \pi/4 $. For $ N = 100 $ the entanglement shows a repeating structure with period $ \pi/2 $, but otherwise no apparent periodicity in between. 
The fluctuations become increasingly at finer timescales as $ N $ grows, where the timescale of the fluctuations occur at
an order of $ \sim 1/N $. The basic behavior can be summarised as follows. The entanglement increases monotonically until a characteristic time $ \tau = 1/\sqrt{2N} $, after which 
very fast fluctuations occur, bounded from above by a ``ceiling'' in the entanglement.   In a realistic BEC with $ N=10^3$-$10^6$ the fluctuations occur even faster than the results of the numerical simulations shown here.

\begin{figure}
\scalebox{0.3}{\includegraphics{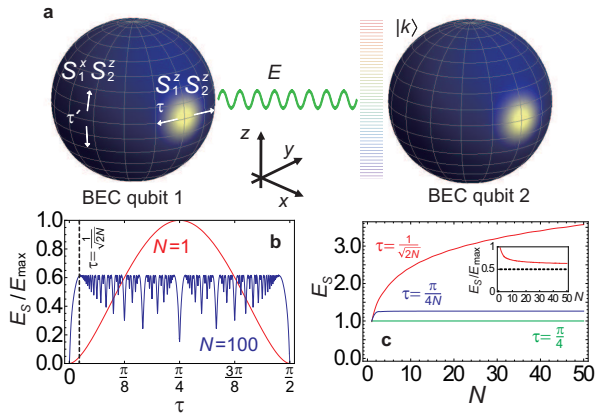}}
\caption{\label{fig1}
Entangling operation between two spinor Bose-Einstein condensates.  (a) A schematic of the operation considered in this paper. The $Q$-functions for the initial spin coherent state $ | \frac{1}{\sqrt{2}},\frac{1}{\sqrt{2}} \rangle \rangle $ on each BEC are shown. The $ S^z S^z $ and  $ S^x S^z $ interactions induces a fanning of coherent states in the direction shown by the arrows.  
(b) The von Neumann entropy normalized to the maximum entanglement ($E_{\mbox{\tiny max}} = \log_2 (N+1) $) between two BEC qubits for  $ N = 1 $ and $ N =100 $ after operation of $ S^z_1 S^z_2 $ for a time $ \tau $. (c) Entanglement at times $  \tau = \frac{\pi}{4N}, \frac{1}{\sqrt{2N}}, \frac{\pi}{4} $ for various boson numbers $ N $. Inset: The same data for $ \tau = \frac{1}{\sqrt{2N}} $, but normalized to $E_{\mbox{\tiny max}} $. }
\end{figure}

\begin{figure}
\scalebox{0.25}{\includegraphics{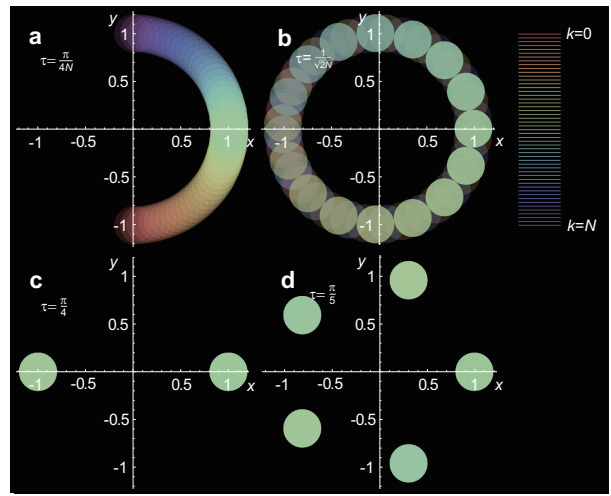}}
\caption{\label{fig2}
Visualization of the entangled state (\ref{entangledstate}) for gate times (a) $ \tau = \frac{\pi}{4 N} $ (b) $ \tau = \frac{1}{\sqrt{2N}} $  (c) $ \tau = \frac{\pi}{4} $ (d) $ \tau = \frac{\pi}{5} $.  Eigenstates of the $ S^z $ operator, $ | k \rangle $ are entangled with spin coherent states of the same color as given in the key.  The radii of the circles correspond
to approximate distances for diminishing overlap of the spin coherent states. Colors give the entangled states given in the key,
and the transparency of the circles reflect the size of the coefficients in (\ref{entangledstate}). All plots correspond to $ N = 50 $.}
\end{figure}

What is the origin of the complex behavior of the entanglement?  Figure \ref{fig1}a shows the 
$ Q $-function \cite{gross12} for the initial state $ \tau = 0 $ of one of the BECs  on the surface of the normalized Bloch sphere. Evolving
the system now in $ \tau $, a visualization of the states is shown in Figure \ref{fig2}. Keeping in mind that the spin coherent states form a quasi-orthogonal complete set of states $ \langle \langle \frac{e^{i \phi'}}{\sqrt{2}}, \frac{e^{-i \phi'}}{\sqrt{2}} | \frac{e^{i \phi}}{\sqrt{2}}, \frac{e^{-i \phi}}{\sqrt{2}} \rangle \rangle \approx 
\exp [ -N (\phi-\phi')^2/8] $, we represent various spin coherent states in the sum (\ref{entangledstate}) each as a circle of radius $ r = \sqrt{\frac{2}{N}} $, which corresponds to the distance where the
overlap between two spin coherent states start to diminish exponentially.  As all spin coherent states in (\ref{entangledstate}) are along the equator of the Bloch sphere $ z=0 $, we flatten the Bloch sphere along the $ z $-direction, such that each spin coherent state is located at an angle $ \phi $ in the $x$-$y$ plane. The color of the 
circle corresponds to the particular $ | k \rangle $ state that the spin coherent state is entangled with (see also Figure \ref{fig1}a).  The opacity of the 
circles are adjusted such that it corresponds to the coefficient in (\ref{entangledstate}), so that smaller terms appear
more transparent.  

There are several characteristic times which we consider separately. As $ \tau $ is increased, the circles fan out in 
both clockwise and counterclockwise directions. At $ \tau = \frac{\pi}{4N} $, the extremal states $ k=0,N  $ reach the $ \pm y $ directions.   For $ N = 1 $ (the standard qubit case) this gives a maximally entangled state. Figure \ref{fig1}c shows the amount of entanglement for times $ \tau = \frac{\pi}{4N} $, which shows that it quickly approaches the asymptotic value of $ E_S \approx 1.26 $. A maximally entangled state would have entanglement $ E_{\mbox{\tiny max}} = \log_2 (N+1) $, and thus gate times of $ \tau = \frac{\pi}{4N} $ correspond to relatively small amounts of entanglement, equivalent to approximately one pair of entangled qubits. 

The next characteristic time is $ \tau = \frac{1}{\sqrt{2N}}$.  This is the time when the spin coherent states are separated far enough such 
that adjacent circles have no overlap (see Figure \ref{fig2}b). The entanglement, as can be seen in Figure \ref{fig1}b, reaches its maximal 
value at this time, and beyond this executes fast fluctuations briefly returning to this value.  It is also a characteristic time due to 
the weight factors in (\ref{entangledstate}).  Approximating  $ \sqrt{\frac{1}{2^N}{N \choose k}} \approx (\frac{2}{\pi N})^{1/4} \exp  [-\frac{1}{N}(k-N/2)^2 ] $, we see that the only the terms between $ k = N/2 \pm \sqrt{N}  $ have a significant weight in the summation and the other terms only contribute and exponentially small amount. At the time $ \tau = \frac{1}{\sqrt{2N}}$, these states are spread out over the unit circle, since the angular positions of the spin coherent states in (\ref{entangledstate}) reach order unity for the first time.  These two observations give us a simple way of estimating the asymptotic value of the entanglement in the limit $ N \rightarrow \infty $. 
Starting from (\ref{entangledstate}) and discarding small weight contributing states, we have $ \frac{1}{\sqrt{2N}} \sum_{k=N/2-\sqrt{N}}^{k=N/2+\sqrt{N}} | \frac{e^{i(N-2k)\tau}}{\sqrt{2}} , \frac{e^{-i(N-2k)\tau}}{\sqrt{2}} \rangle \rangle_1 | k \rangle_2
$ where we have approximated the binomial factor within the range to be constant.  Assuming that the spin coherent states are orthogonal, this
gives an entanglement of $ E_S \approx \log_2 \sqrt{N} $.  Relative to the maximal value this gives 
\begin{align}
\lim_{N\rightarrow \infty} E_S/ E_{\mbox{\tiny max}} = 1/2  .
\label{asymptoticent}
\end{align}
This result is interesting since it shows that even in the limit of $ N \rightarrow \infty $ entanglement survives, on a macroscopic scale. 
It is clear from this example that classical transition does not occur simply due to the large number of particles in the system, but only when decoherence is also present. The inset of Figure \ref{fig1}c shows results consistent with the asymptotic result (\ref{asymptoticent}).

\paragraph{Devil's crevasse}

We now consider the origin of the fast fluctuations in the entanglement in Figure \ref{fig1}b.  Due to the periodicity of the spin coherent states, the circles align with high symmetry at particular $ \tau $ times as seen in Figures \ref{fig2}c and \ref{fig2}d.  These points correspond to sharp dips in the entanglement as seen in Figure \ref{fig1}b. For example, at $ \tau = \pi/4 $, (\ref{entangledstate}) can be 
written
\begin{align}
& \frac{1}{2} \left( | \frac{1}{\sqrt{2}} , \frac{1}{\sqrt{2}} \rangle \rangle_1  + | \frac{1}{\sqrt{2}} , -\frac{1}{\sqrt{2}} \rangle \rangle_1  \right) | \frac{e^{i \pi N/4}}{\sqrt{2}} , \frac{e^{-i \pi N/4}}{\sqrt{2}} \rangle \rangle_2 \nonumber \\
& + \frac{1}{2} \left( | \frac{1}{\sqrt{2}} , \frac{1}{\sqrt{2}} \rangle \rangle_1  - | \frac{1}{\sqrt{2}} , -\frac{1}{\sqrt{2}} \rangle \rangle_1  \right) | -\frac{e^{i \pi N/4}}{\sqrt{2}} , \frac{e^{-i \pi N/4}}{\sqrt{2}} \rangle \rangle_2 .
\label{schrodingercat}
\end{align}
For large $ N $, the states $ | \frac{1}{\sqrt{2}} , \frac{1}{\sqrt{2}} \rangle \rangle $ and $ | \frac{1}{\sqrt{2}} , - \frac{1}{\sqrt{2}} \rangle \rangle $ require flipping $ N $ bosons from $ +x $ to $ -x $ eigenstates, which is a Schrodinger cat state. Taking into account that 
$ \langle \langle \frac{e^{i\phi}}{\sqrt{2}}, -\frac{e^{i\phi}}{\sqrt{2}} | \frac{e^{i\phi}}{\sqrt{2}}, \frac{e^{i\phi}}{\sqrt{2}} \rangle \rangle = 0 $ for any $ N $, (\ref{schrodingercat}) is precisely equivalent to a Bell state, which has an 
entanglement $ E_S = \log_2 2 = 1 $ for all $ N $, in agreement with the numerical results of Figure \ref{fig1}c. 

Similar arguments may be made for any $\tau$  that is a rational multiple of $ \pi/4 $. Consider a general time $ \tau = \frac{m \pi}{4d} $, where $ m,d $ are integers. The angular difference between adjacent circles in Figure \ref{fig2} may be deduced to be $ \Delta \phi = 4 \tau = \frac{m\pi}{d} $. The number of circles must then be the first integer multiple of $ \Delta \phi $ that gives a multiple of $ 2 \pi $.  This is $  \mbox{LCM} (m/d,2) d /m $, where $ \mbox{LCM} $ is the least common multiple.  The amount of entanglement at such times can then be written
\begin{align}
E_S = \log_2 \left[ \frac{ \mbox{LCM} (m/d,2) d }{m} \right] .
\label{rationalent}
\end{align}
The entanglement at times that are an irrational multiple of $ \pi/4 $ do not have any particular alignment of the circles, 
and thus can be considered to be approximately randomly arranged over $ \phi $.  Similar arguments to (\ref{asymptoticent}) may then 
be applied, and at these points the entanglement jumps back up to $ E_S/E_{\mbox{\tiny max}} = 1/2 $ for $ N \rightarrow \infty $. 

Since the set of rational numbers is everywhere dense, in the limit of $ N \rightarrow \infty $, the entanglement has a pathological behavior where there are an infinite number of
dips between any two values of $ \tau $.  Such a dependence of whether a parameter is either rational or irrational is familiar in physics contexts from models such as the long-range antiferromagnetic Ising model \cite{bak82}, where the filling factors equal to a rational number occupy a finite extent of the 
chemical potential, forming a devil's staircase.  In analogy to this, 
we may call the behavior of the entanglement a {\it devil's crevasse}, where every rational multiple of $ \pi/4 $
in the gate time gives a sharp dip in the entanglement.  It is interesting that unlike the devil's staircase which has origins in two competing periodicities forming either commensurate or incommensurate phases 
\cite{bak82b}, here the only periodicity is that of the Bloch sphere, and the mechanism of the pathological behavior is different. 

\begin{figure}
\scalebox{0.3}{\includegraphics{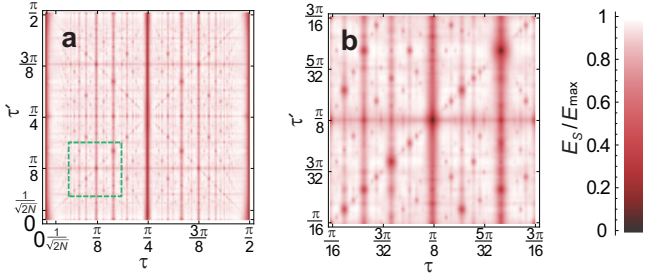}}
\caption{\label{fig3}
Entanglement under the concatenated operations of $ S^z_1 S^z_2 $ and $ S^x_1 S^z_2 $ for times $ \tau $ and $ \tau' $ 
respectively. Plots are for the (a) first period $ \tau, \tau' = [0,\pi/2] $ beyond which the structure repeats, (b) 
a zoomed in region showing the extension of the fine structure to intervals $ \Delta \tau, \Delta \tau' \sim \frac{1}{N} $. 
$ N = 100 $ is used for both plots. }
\end{figure}

Armed with this interpretation, it is natural to extend the operation to not only rotations in the $x$-$y$ plane, 
but the full Bloch sphere.  Namely we may first perform a $ S^z_1 S^z_2 $ operation, then perform another entangling gate 
around another axis, say the 
$x$-axis (see Figure \ref{fig1}a) to perform the operation $ e^{-i S^x_1 S^z_2 \tau}  e^{-i S^z_1 S^z_2 \tau} | \frac{1}{\sqrt{2}}, \frac{1}{\sqrt{2}} \rangle \rangle_1 
| \frac{1}{\sqrt{2}}, \frac{1}{\sqrt{2}} \rangle \rangle_2 $.  The entanglement in the two parameter space of $ (\tau,\tau') $ is shown in Figure \ref{fig3}. We see a highly detailed
fractal-like structure, with sharp dips in the entanglement at rational multiples of $ \pi/4 $ for both $ \tau $ and 
$ \tau' $.   The fine detail is beyond the resolution of the graph 
shown in Figure \ref{fig3}a, as can be seen by a zoomed in plot shown in Figure \ref{fig3}b.  The detail extends to angular
intervals of $ \Delta \tau, \Delta \tau' \sim \frac{1}{N} $, thus in the limit of $ N \rightarrow \infty$ the fine
structure extends to arbitarily small regions of the graph.

\paragraph{Decoherence}

In any realistic system decoherence is present, and considering the large number of particles that are typically present 
in a BEC, a question remains of how observable such entanglement is. As a generic model of decoherence we consider 
Markovian dephasing to occur on each of the spin states 
\begin{align}
\frac{d \rho}{d \tau} = -i [ S^z_1 S^z_2 , \rho] -\frac{\Gamma_j}{2} \sum_{n=1}^2 
\left[ (S^j_n)^2 \rho - 2 S^j_n \rho S^j_n + \rho (S^j_n)^2 \right] ,
\label{dephasingmaster}
\end{align}
where we consider the effects of $ j=z,x $ separately. We find that there is a strong dependence on how well the entanglement survives
depending on the characteristic gate times.  This is summarized by the behavior
of the ratio between the entanglement, as measured by the logarithmic negativity \cite{vidal02,plenio05}, with and without decoherence.  For 
the $z $-decoherence, we find a power law decay of the entanglement as $ \frac{E_{\mbox{\tiny neg}} (\Gamma_z)}{E_{\mbox{\tiny neg}} (\Gamma_z=0)} \sim N^{-\gamma} $, where $ 0\le\gamma\le 1 $.  For $x $-decoherence, there is a strong dependence on the time of evolution
\begin{align}
\frac{E_{\mbox{\tiny neg}} (\Gamma_x)}{E_{\mbox{\tiny neg}} (\Gamma_x=0)} \sim \left\{
\begin{array}{ll}
\mbox{const.} & \tau \sim 1/N\\
N^{-\gamma} & \tau \sim 1/\sqrt{N}\\
e^{-N^2} & \tau \sim O(1)\\
\end{array}
\right.
\label{asymptoticvalues}
\end{align}
For short gate times of $ \tau \sim 1/N $ the entanglement is very robust even with decoherence.  For longer gate times $  \tau \sim O(1) $, 
the entangled states are rather fragile, due to the appearance of Schrodinger cat-like states such as (\ref{schrodingercat}), with the border between these two behaviors is $ \tau \sim 1/\sqrt{N} $.  We thus anticipate that in practice gate times up to $ \tau \le  1/\sqrt{N} $ should be robust enough to observable experimentally.  However, in analogy to a decoherence free subspace approach \cite{lidar03}, by arranging the interaction to be in a commuting basis to the decoherence (such as in the $ z $-decoherence case here), states with $ \tau > 1/\sqrt{N} $ may be prepared.

\paragraph{Experimental detection}

In order to experimentally detect the type of entanglement that is discussed in this paper, it is beneficial to 
have simple quantities that serve as signatures of entanglement between two BECs.  For small times $ \tau \sim 1/N $,
and initial states with both BEC qubits in maximal $ x $-eigenstates we may approximate the spins $ S^y $ and $ S^z $
as conjugate position and momentum variables \cite{braunstein05} according to $ X_i = \frac{S^y_i}{2 \sqrt{ \langle S^x_i \rangle}} $, 
$ P_i = \frac{S^z_i}{2 \sqrt{ \langle S^x_i \rangle}} $, where $ i = 1,2 $. Following the arguments of Ref. \cite{duan00}, for a general separable state we derive the inequality
\begin{align}
\langle [ \Delta (S^y_1 - S^z_2)]^2 \rangle + \langle [ \Delta (S^y_2 - S^z_1)]^2 \rangle \ge 2 \langle S^x_1 \rangle + 2 \langle S^x_2 \rangle .
\label{duaninequality}
\end{align}
This inequality may be used to test for the presence of entanglement generated by the $ S^z_1 S^z_2 $ interaction.  To 
see why it is this combination of operators which signals the presence of entanglement, let us explicitly derive 
this quantity under the assumption $ \langle S^x_i \rangle =N $. the entangling Hamiltonian in these variables is $ H = 4JN P_1 P_2 $, and we may derive the 
Heisenberg equations of motion
\begin{align}
X_1(\tau) &= X_1(0) +2N \tau P_2(0) \nonumber \\
X_2(\tau) &= X_2(0) +2N \tau P_1(0) \nonumber \\
P_i(\tau) & = P_i(0) .  \label{heisenberg}
\end{align}
Although the dynamics of these operators is rather different to two-mode squeezing due to the different Hamiltonian considered here, it is still possible to form EPR-like quantities with suppressed noise fluctuations.  Using (\ref{heisenberg}) we derive for an initially uncorrelated 
state
\begin{align}
\langle [ \Delta (S^y_1(\tau) - S^z_2(\tau))]^2 \rangle & = \langle [ \Delta (S^y_2(\tau) - S^z_1(\tau))]^2 \rangle  \nonumber \\
 & = N(1-2N\tau)^2+N . \label{eprevolution}
\end{align}
It is apparent that the above expression may violate (\ref{duaninequality}) which unambiguously shows the presence
of entanglement between the BECs.  We note that beyond times of order $ \tau \sim 1/N $, the continuous variables approximation breaks
down, so \ref{eprevolution}) is only valid in this region.  We may understand this as
times where the entanglement creates distributions of coherent states that are no longer on one hemisphere of the 
Bloch sphere (see Figure \ref{fig2}a).  Heuristically, the approximation breaks down due to the position and momentum variables assuming a ``flat'' underlying topology, as opposed to the ``spherical'' topology for spins. Such oversqueezing 
for two-mode squeezing is also discussed in Ref. \cite{riedel10}.

\paragraph{Conclusions}

We have shown that interactions of the form $ S^z_1 S^z_2 $ between two BECs exhibit a rich fractal-like behavior in terms of the 
entanglement generated.  There is a remarkable dependence of the entanglement upon whether the gate times are a rational or irrational 
multiple of $ \pi/4 $, where sharp dips in the entanglement form.  While typically entanglement between macroscopic objects are highly
suppressed due to accelerated decoherence, for gate times $ \tau \le 1/\sqrt{N} $ these should be robust even in the 
presence of decoherence.  For the observation of entaglement beyond these times, arranging for interaction to be in a commuting basis to the 
decoherence would seem to be necessary.  We note that while entanglement between macroscopic objects have been observed before \cite{sherson06,julsgaard01,chou05,bao12}, it is 
always in the small entanglement limit of the order of unity.  The 
present case goes beyond this as it is truly macroscopic entanglement {\it i.e.} the entanglement is of the 
order of the maximal entanglement $ E \sim E_{\mbox{\tiny max}} /2 $. It is interesting that for quantum information purposes, typically only short gate times should
be necessary \cite{byrnes12}, which allow for the possibility that such systems could be used as viable quantum memories.  The large
duplication of the quantum information in the spin coherent state (\ref{singlequbitstate}) would then allow for a robust
way of storing and manipulating the data.  

T. B. thanks Alexey Pyrkov and Ebubechukwu Ilo-Okeke for discussions.  This work is supported by the Transdisciplinary Research Integration Center, the Okawa foundation.



\begin{thebibliography}{99}

\bibitem{bohi09} P. B{\"o}hi, M. F. Riedel,  J. Hoffrogge, J. Reichel, T. W. H{\"a}nsch, P. Treutlein, Nature Phys. {\bf 5}, 592 (2009). 

\bibitem{treutlein06} P. Treutlein, T. Steinmetz, Y. Colombe, B. Lev, P. Hommelhoff,  J. Reichel, M. Greiner,  O. Mandel, A. Widera, T. Rom, I. Bloch, T. W.  H{\"a}nsch,  Fortschr. Phys. {\bf 54}, 702 (2006). 

\bibitem{riedel10}
M. F. Riedel,  P. B{\"o}hi, Y. Li,  T. W. H{\"a}nsch, A. Sinatra, P. Treutlein, Nature {\bf 464}, 1170 (2010). 

\bibitem{gross12} C. Gross, J. Phys. B: At. Mol. Opt. Phys. {\bf 45}, 103001 (2012).

\bibitem{sorensen00}
A. S{\o}rensen,  L.-M. Duan, J. I. Cirac, P. Zoller, Nature {\bf 409}, 63 (2000). 


\bibitem{toth09}
G. Toth, C. Knapp, O. G{\"u}hne, H. J. Briegel, {\it Phys. Rev. A} {\bf 79}, 042334 (2009). 

\bibitem{lettner11}
M. Lettner, {\it et al.} Phys. Rev. Lett. {\bf 106}, 210503 (2011). 


\bibitem{colombe07} Y. Colombe,  T. Steinmetz,  G. Dubois, F. Linke, D. Hunger, J. Reichel,  Nature {\bf 450}, 272 (2007). 

\bibitem{byrnes12} T. Byrnes, K. Wen, Y.  Yamamoto, Phys. Rev. A {\bf 85}, 040306(R) (2012).

\bibitem{byrnes12b}
T. Byrnes, World Academy of Science, Engineering and Technology {\bf 63}, 542 (2012).


\bibitem{pyrkov13} A. Pyrkov and T. Byrnes, arxiv: 1305.2479. 


\bibitem{treutlein06b} P. Treutlein, T. W. H{\"a}nsch, J. Reichel, A. Negretti,  M. A. Cirone, T. Calarco, Phys. Rev. A  {\bf 74}, 022312
(2006).

\bibitem{milburn86} G. J. Milburn, Phys. Rev. A {\bf 33}, 674 (1986).

\bibitem{stobinska08} M. Stobi{\'n}ska, G. J. Milburn,  K. W{\'o}dkiewicz, Phys. Rev. A {\bf 78}, 013810 (2008). 

\bibitem{vidal02} G. Vidal,  and R. F. Werner, Phys. Rev. A {\bf 65}, 032314 (2002).  

\bibitem{plenio05}  M. B. Plenio, Phys. Rev. Lett. {\bf 95}, 090503 (2005).

\bibitem{bak82} P.  Bak and R. Bruinsma, Phys. Rev. Lett {\bf 49}, 249 (1982). 

\bibitem{bak82b} P. Bak, Rep. Prog. Phys. {\bf 45}, 587 (1982).

\bibitem{sherson06}
J. F. Sherson  {\it et al.}, Nature {\bf 443}, 557 (2006).

\bibitem{julsgaard01}
B. Julsgaard,  A. Kozhekin, E. S. Polzik, Nature {\bf 413}, 400 (2001).

\bibitem{chou05}
C.W. Chou {\it et al.}, Nature {\bf 438}, 828 (2005).

\bibitem{bao12}
X.H. Bao {\it et al.}, {\it Proc. Nat. Acad. Sci.} 10.1073/pnas.1207329109  (2012).

\bibitem{lidar03}
D. A. Lidar, K. B.  Whaley, Irreversible Quantum Dynamics pp. 83-120  (Springer Lecture Notes in Physics vol. 622, Berlin, 2003).  

\bibitem{braunstein05}
S. Braunstein and P.  van Loock,  Rev. Mod Phys. {\bf 77}, 513 (2005). 

\bibitem{duan00}
L. M. Duan,  G. Giedke,  J. I. Cirac, and P. Zoller,  Phys. Rev. Lett. {\bf 84}, 2722 (2000)




\end{thebibliography}
\end{document}